# An ultrahigh-speed digitizer for the Harvard College Observatory astronomical plates

R.J. Simcoe [ξ,1], J.E. Grindlay[1], E.J. Los[1], A.Doane[1], S.G. Laycock[1], D.J.Mink[2], G. Champine[1], A.Sliski[1]

[1] Harvard College Observatory, 60 Garden Street, Cambridge, MA 02138
[2] Smithsonian Astrophysical Observatory, 60 Garden Street, Cambridge, MA 02138

**ABSTRACT**

A machine capable of digitizing two 8 inch by 10 inch (203 mm by 254 mm) glass astrophotographic plates or a single 14 inch by 17 inch (356 mm by 432 mm) plate at a resolution of 11 μm per pixel or 2309 dots per inch (dpi) in 92 seconds is described. The purpose of the machine is to digitize the ~500,000 plate collection of the Harvard College Observatory in a five-year time frame. The digitization must meet the requirements for scientific work in astrometry, photometry, and archival preservation of the plates. This paper describes the requirements for and the design of the subsystems of the machine that was developed specifically for this task.

## 1. INTRODUCTION –GOALS AND HISTORY

The Harvard College Observatory (HCO) maintains a collection of more than 500,000 glass astrophotographic plates that cover both the northern and southern skies from the 1880s to the 1980s. The collection also contains daguerreotypes and plates from 1849 to the 1880s that are primarily of historic interest. The main body of the collection represents more than 25% of the world's total of stored and cataloged wide-field astrophotographs, by far the world's largest and most complete collection in both time and sky coverage. However, in its current analog form this large database is much under-utilized because of the difficulties of needing physical access at Harvard, special skills, and much manual labor to search through the catalog, the collection, and then examine the plates by eye. The goal of this project is to show the feasibility of converting this analog database to digital form to make it accessible to the computerized tools of modern astronomy.

One of us (JEG, the project's Principal Investigator), has been interested in digitizing the massive Harvard astronomical plate collection since the 1980s since it provides a unique resource, barely touched, for the long-term variability of compact objects and quasars that in turn enable, for example, measures of black hole mass. However, only recently has technology (for rapid imaging, scanning, data transfer, and storage) become available to make such a massive digitization project conceivable. In 2001 Grindlay encouraged A. Doane, curator of the plate collection, to look at the feasibility of digitizing the plates, mostly 8 inch by 10 inch, with commercial flat bed scanners. Doane worked with Grindlay and D. Mink to test several of the best available commercial scanners and found that good photometric science with the plates was possible, but that it took ~20 minutes to scan a single 8 inch by 10 inch plate. Such long scan times as well as reliability issues made commercial scanners unsuitable for a collection of 500,000 plates. Given the long history of association and cooperation between the professional and the amateur astronomy community at Harvard, Doane gave a talk on this work at a meeting of the Amateur Telescope Makers of Boston (ATMoB), which meets monthly at the HCO. One us (RJS, an ATMoB member), believed that this problem could be tackled with a specialized digitizing machine and volunteered to help develop one. Accordingly, he began working with Grindlay to investigate the feasibility of building such a machine with today's technology. This culminated in a proposal to NSF to design and build a machine to demonstrate that it was feasible to both digitize the collection and store it electronically online. This

---

[ξ] rjsimcoe@cfa.harvard.edu; phone :1 617 495 3362

was the genesis of the "Digital Access to a Sky Century at Harvard" (DASCH) project. The overall project and science goals, together with initial results on calibration scans of an open cluster (M44) are given by Grindlay et al (in preparation). The initial tools for astrometry and photometry of the digitized images are described by Laycock et al (in preparation).

There have been a small number of machines built around the world with the specific purpose of digitizing astrophotographs. Of the machines in the USA, the last one built was designed over 20 years ago and was optimized for astrometry. Called the PMM[1, 2], this machine, at the Naval Observatory in Flagstaff, took 1-4 hours to scan (13.6 µm pixels, 8 bit digitization) and process a plate and was recently decommissioned. A machine called StarScan[3] at the Naval Observatory in Washington DC, first built in the 1970s and updated in the 1990s, is still operational. These machines used CCD area arrays.

For photometry the standard workhorse was the PDS developed in the early 1970s and available into the 1990s. It is an x-y table fitted with servomotors to drive the stages and an iris to measure one star diameter at a time. By using a fixed iris, the PDS can be turned into a spot scanner that takes from 4 hours to as much as 24 hours (depending on size) to scan a plate. In the 1980s there were several GAMMA machines based on the PDS at the Space Telescope Science Institute (STScI) that were used to produce the guide star catalog for Hubble. Yet another adapted PDS machine, the significant APS[4] machine at the University of Minnesota, was used for a sky survey there.

In the early 1990s a SuperCOSMOS[5] machine was developed in England. It used a linear photo sensor array and took less than 2 hours to digitize a plate.

A new machine called D4A[3] is being developed in Belgium. It is designed to do glass astrophotographic plates and aerial photographic film. It uses an area array and will digitize an area 350 cm (13.75 inches) on each side and is designed for a throughput of six photographs 240 mm (9.45 inches) on each side per hour.

These machines were generally built or adapted to do astrometry measurements on wide field Schmidt telescope plates from telescopes around the world. The Schmidt plates were mostly 350 mm square. A full sky survey, using these plates, typically required digitizing at most a few thousand plates.

All of these machines provided a repeatability of measured centroid locations to about 1µm. They provided a photo density (see section 4.4.4) measurement range of about 2-3$D_p$ (8-10 bits), stellar coordinate accuracy in the range of .2 arcsec on the Schmidt plates, and photographic magnitude repeatability in the 0.2 magnitude range[5]

The International Astronomical Union (IAU), since 2000, has had a working group on the Preservation and Digitization of Photographic Plates (PDPP). The focus of the group has been advisory rather than practical, though some members have used commercial flat bed scanners to digitize several small and one large collection of generally smaller (5 by 5 inch or 5 by 9 inch) plates. But, so far, little work has been reported on the accuracy and usefulness of the digitized data for astrometric or photometric measurements from these scanners.

A history of measuring machines is available on the web site http://www.astro.virginia.edu/~rjp0i/museum/index.html.

## 2. SYSTEM REQUIREMENTS

For the extensive Harvard collection we believed we needed to have a digitizer accurate enough to meet the astrometric, photometric, and archival needs of the astronomy community and also fast enough to make the project feasible. In addition the digitizer needed to handle a wide variety of plate sizes from many different telescopes and eras.

### 2.1 Digitizing Time

A prime requirement we set was to be able to digitize the whole collection in a five-year timeframe once production started. This meant we would have to digitize ~100,000 plates per year or about 400 plates per day. To accomplish this we would need to digitize a plate about every minute on average, and we assumed that perhaps half of that time would be needed to simply handle a plate as it was placed on and taken off of the digitizer. The minute per plate goal required the new digitizer to be ~100 – 200 times faster than previous special purpose machines. Fortunately, technology advances have made it feasible to build such a machine.

## 2.2 Astrometric

Astrometric measurements are used to determine the positions of stars in relation to a World Coordinate System (WCS). The positions and proper motions of many stars are cataloged, at certain epochs in time. The long time-base and extensive coverage, with on the order of 1000 plates of any particular star of brightness up to $15^{th}$ magnitude and 100-500 plates of stars of $15^{th}$-$18^{th}$ magnitude, are unique properties of the HCO plate collection. By measuring (at different points in time) the locations of stars that are suspected of moving or stars that have not been cataloged against the locations of stars that are believed to be relatively motionless, we can better predict a star's future location. Although the Harvard digitizer is not primarily an astrometric measuring engine, it will provide unique long-term measurements of previously unknown high proper motion objects (e.g. undiscovered asteroids; and high velocity stars).

The digitizer provides a measurement grid that can be used to measure the relative positions of the stars on the plates. The primary ruler is the CCD 11 μm pixel pattern. This "ruled grid" must then be moved accurately in both x and y directions to create a mosaic grid, accurately registered (with positional uncertainties in the positions of the tiled pixel array ultimately <<11 μm) over the whole plate. Even though the grid is formed by the 11 μm pixels of the CCD, star positions can be interpolated to ≤1 μm by computing the position of their centroids. To make accurate measurements over the whole plate, the mechanism for moving the plate with respect to the CCD needs to be accurate to about 5% of a pixel or about 0.5 μm.

## 2.3 Photometric

The most important and unique scientific work using the plates are time variability studies. The long historic time-base of the collection makes it particularly well suited for this work. We are developing an automated magnitude measurement process because it must be computed in a special way, with local calibrations, for each plate. Photometry of variable objects, from stars to quasars, over the uniquely long time frame and densely sampled coverage provided by the HCO plates, will enable DASCH to extend backwards in time the powerful new time variability projects and full-sky surveys now being conducted or planned using modern telescopes and digital detectors.

Photometric measurements are made by comparing the area and photographic density of the exposed images of stars (or other objects) on the plate to the corresponding values for stars of known brightness or astronomical magnitude in a prescribed optical passband. The star image diameters (isophotal diameters, above a set threshold) range from about 30μm for the dimmest star images to many hundreds of μm for the bright stars. Since each plate can have unique exposure times, differences in seeing conditions, photographic emulsions, and developing conditions, each plate must be calibrated against known stars of stable brightness and colors on that plate. To calibrate a plate, we must first identify all of the stars on it against catalogs to find calibrators. This was one of the impediments to doing DASCH earlier. Only very recently have the highly accurate (~0.02-.05 magnitude) catalogs, including color data (magnitudes in several passbands) over much of the sky (and soon full-sky) in digital form become available. This catalogue information is needed to correct for the reddening due to the airmass for the exposure and for the effective passband recorded by each plate. Details of our photometric analysis system are provided by Laycock et al (in preparation).

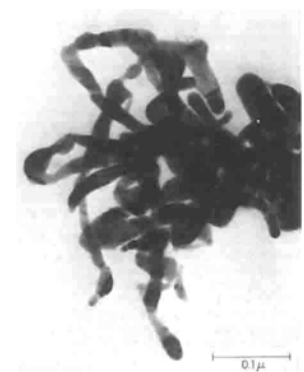

**Figure 1** Developed crystal grain turns to filaments

## 2.4 Archival

To meet archival requirements, we must capture all of the information that is on the plate. There should be little reason to ever physically re-examine the plate. Once digitized, plates can then be placed in a long term storage area where physical access is still possible but not as easily as in the current plate stack.

One of the first questions to arise when setting requirements for scanning a plate for archival purposes is "What size pixels are needed to capture all of the information on a photographic plate?" There is a general awareness that photographic emulsions contain silver halide crystals that become chemically activated by photons, allowing them to turn into silver when developed. What is less well known is that as the crystalline grains are developed, they turn into silver filaments. "A very small silver halide grain can be developed to yield only one filament per grain … a large grain commonly forms a mass of many filaments that roughly resembles a wad of steel wool in structure."[6] Figure 1 shows a very high magnification microphotograph of a grain developed into silver filaments. Figure 2 shows different halides form similar but different `filamentary tangles[7]. Understanding how the tangled filamentary structures that make up the star images on the plate relate to the pixel sizes of a CCD imager requires more complex analysis. Traditionally, film-based systems have been defined in terms of resolution or modulation transfer function (MTF). The problem then is to equate a film MTF to an equivalent MTF for square pixels of a given size on a CCD imaging sensor. The one dimensional MTF for a sensor to record images with spatial frequency $\nu$ is given by: MTF $(\nu)$ = $\sin(\pi D\nu)/\pi D\nu$. If we normalize the measured film MTF and the ideal sensor MTF at the 50% point, then one can relate the size of the imaging aperture of size D to the 50% response frequency $\nu_o$, by the equation: $\pi D\nu_o$ = 1.9. Using the measured MTF values for film, this yields the effective pixel aperture D for each film. Table 1 shows the effective pixel size for common film speeds[8, 9]. Old astronomical plates are not well characterized, but from this analysis a pixel size in the 10-11 µm region will capture the information on even the best of the old plates. Measurements on plates from POSS II have also indicated that scanning pixel sizes of 15µm and below capture all the information on modern plates.[10]

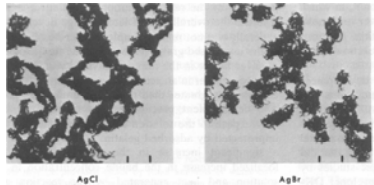

**Figure 2** Developed filament tangles of halides AgCl and AgBr

| ISO Speed | Pixel Size |
|---|---|
| 100 | 12 µm |
| 200 | 14 µm |
| 400 | 19 µm |
| 1000 | 26 µm |

**Table 1** Film Equivalent Pixel Sizes

The pixel size also greatly affects digital storage requirements. An 8 by 10 inch plate digitized with 11 µm pixels requires 740 Mb of storage. If the pixel size is halved, the storage requirement quadruples. The data needs to be stored in its raw form for archival purposes. The raw data in the form of tiled CCD images need to be processed to create a digital mosaic of the plate which is the primary image for deriving the WCS astrometric solution and photometric catalog for objects on the plate. To record archival historical notations, often made on the glass side of the plate, we take a high-quality digital image (with a commercial digital camera) of that side of the original plate before cleaning in preparation for the digital scan. Additional images of the original logbook entry and the covering jacket associated with each plate also must to be taken for archiving. The end result is to have all of the information about and on the plate in electronic form. We estimate that for each 8 by 10 inch astrophotograph, we will need somewhat more than 2 GB of storage. The whole project will ultimately require 1-1.5 PB of storage with about 500TB of that online, an amount that has just recently become practical to implement.

## 2.5 Physical constraints:

The design of the digitizer was physically constrained to have assembled subsystems that could be brought into the bottom floor of the HCO plate stacks. The plate stacks are in a building that was designed and constructed in the 1920s to be an earthquake proof facility. The exterior walls are 12-18 inch thick masonry. The plates are stored in a skeletal steel support grid that is physically separate from the external walls. The steel grid has a support post about every 55 inches on a rectangular grid. The metal cabinets that store the plates normally hide these posts, which are necessary to support the estimated 160 tons of glass plates in the collection. In order to make a work area for the digitizer, and have a place to prepare and clean the plates (by creating 2 lab rooms on the ground floor of the plate stacks), we first had to co-ordinate a move and compaction of 100,000 plates, fully 20% of the collection, between the three floors of the stack area.

The only access to the area, other than very narrow circular staircases, is through a window that has an opening about 4 feet by 4 feet. The size of the window plus the grid of steel supports severely limited how large the table could be. In addition, the largest plates, which are 14 by 17 inches, set a minimum size for the table. Fortunately we were just able to squeeze our digitizer into this area. We constructed a 3D model of both the machine and the room to be sure that we did not have problems when everything was ready for assembly and integration.

Many previous purpose-built scanners are in tightly climate controlled rooms. Because the digitizing process for a plate took several hours or more, it was necessary to keep the temperature stable for that time period so that the plates would not change dimensions during scanning. We do not have such careful control of the climate in the digitizing room at Harvard, but because the digitizing times are so short, the time constants for dimensional change of the glass are much greater than the time the plates are on the table actually being digitized. In addition, both the astrometry and photometry are "self-calibrated" by using the several thousand stars on a typical plate. However we are very careful to maintain long-term absolute calibrations of the table, camera, and light system. Flat field data and linearity data is taken before every digitizing run and an additional set of linearity data is taken immediately after that run.

## 3. DIGITIZER TECHNOLOGY

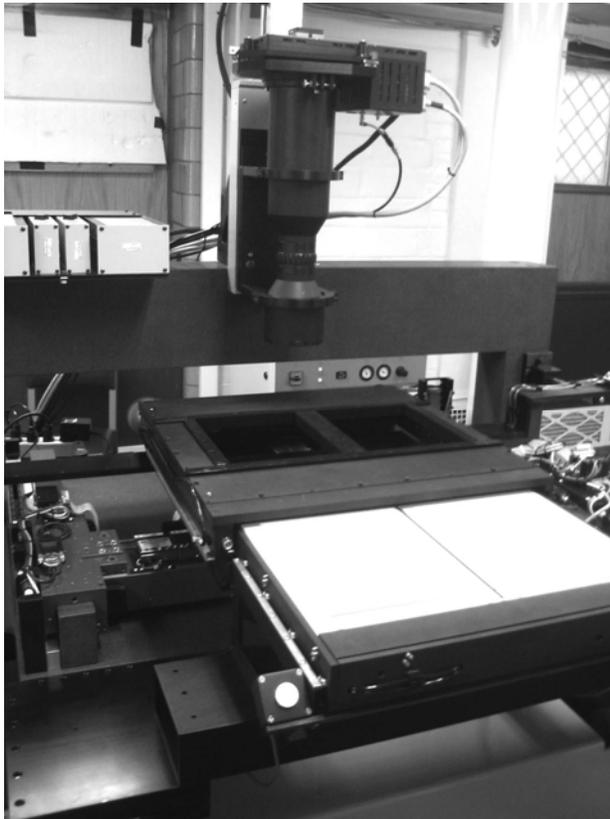

**Figure 3** Digitizer with Fixture Tray Extended

Our design is very similar to previous application-specific digitizers, but we expected that by using up-to-date precision x-y tables (using air bearings and linear motors), new large area CCD cameras, and fast computers with the latest I/O systems we could provide the speed and accuracy needed to meet the our requirements. Performance goals, however, would require many parts of the system to be at the cutting edge of technology.

A commercial scanner typically uses a linear array of photo-sensors and takes mechanically discrete pixel length steps to scan the length of a plate. In order to collect enough light for the pixel sensors, it must wait 8-10msec between mechanical steps. High resolution requires small steps, each with a wait period, which is why it takes a long time to scan. The DASCH digitizer uses a custom-made digital camera, with an area sensor, to take a series of frames (pictures) of the plate which are then stitched together in a mosaic to create an image of the whole plate. With an area array, it also typically takes 8 ms per exposure per pixel, but the array can then be mechanically stepped several thousands of pixels. This speeds up the digitizing process by that factor.

### 3.1 CCD – large and fast

A key to building a fast digitizer was to have a large, fast sensor. A large CCD reduces the number of steps that the table must take and a fast one minimizes the time to read out the results of each frame. Most large CCDs were developed for astronomy and are designed for long integration times and slow readout times. Searching for a large, fast sensor, we identified a CCD from ATMEL (San Jose, CA) that had 4k by 4k pixels and could read out 7 frames per second. The pixels, 11 μm square, have full well capacities of 125Ke, and are able to support 12 bit digitization. Even though no camera using this chip was commercially available, we were fortunate enough to have a camera (Model # SI16M8) donated to the project from a company (Photon Dynamics, Inc.) that developed it for its own use. The camera digitizes to 12 bits, can support the read-out of 7 frames a second, and temperature controls the CCD chip to 18 degrees C. The camera does not have a shutter and has a non-standard lens mount. The CCD area is 45 mm on a side, resulting in a 61 mm diagonal. The camera came with no software so one of us (EJL, an ATMoB member), has developed extensive camera control and scanner table integration software for our application (see section 3.6).

### 3.2 Lens

Early searches for a suitable telecentric lens for the large CCD did not turn up any commercially available options. Just before committing to a custom design, we discovered a source (Sill Optics, Wendelstein, Germany) that had a lens, designed for an 8K linear array, with specifications that seemed adequate. The lens has a working distance of 180 mm and a specified working image size of 72 mm. They also indicated a very low distortion of 0.01%, which is a very critical factor. Even 0.01% distortion amounts to ~1/3 of a pixel or about 4 μm. Lens distortion is one of the largest sources of error in the system but has been carefully measured (see section 4.4.1) and accounted for in our scanning pattern.

### 3.3 Air bearing table

Selecting an air bearing table was difficult because of the size of the large plates, the limitations on the size of the table, and the projected weight of the plate holder. We settled on a customized ABL9000 air bearing, linear motor table from Aerotech, Inc. (Pittsburgh, PA). The table was customized to have a range of motion of 460 mm in X and 380 mm in Y. The ABL9000 stages have an H configuration with the center bar of the H being the X axis. The chuck on the X axis, to which the plate holder attaches, rides directly on an air bearing to the granite table surface. This allows supporting the considerable weight of the plate holder very stiffly.

A consequence of this configuration was that the light system (an LED array, which illuminates the glass plate from below; see section 3.5) either had to be part of the plate holder and therefore illuminate a very large area uniformly or it had to be very compact and the plate holder would need to move around a stationary light system. We chose to do the latter and so constrained the design of the light system to be mounted inside a 3 inch wide, one half inch high hollow bar that runs across the table. The plate holder has a slot in it that the light bar runs through, allowing the plate holder to travel the full distance in X and Y without colliding with the light bar. There is 1/8 inch clearance above and below the light bar. Our specifications called for the table to move a 40 lb load (the plate holder and the plate(s); see section 3.4) 25 mm in 300 ms to an accuracy of .2 μm. We chose to use Zerodur® glass scales rather than laser interferometers for the system servo feedback. The glass scales have 4 μm markings which are multiplied by 250 to achieve a servo step size of 10 nm. The table was calibrated with a laser interferometer at the factory. After 2D compensation from the laser calibration we found the absolute accuracy of the table to be 0.2 μm over the working surface. Testing at the vendor showed that the table would perform to spec with a 50 lb (22.7 kg) load.

The table is supported on a custom-designed frame with a compressed-air leveling and vibration-dampening system. This was custom designed to fit within the building constraints. The granite table weighs about 2200 lbs (1000 kg). To have the center of gravity of the table below the contact points of the air isolators, we added aluminum extensions to the front and back of the table. This also gave greater leverage for the isolators while still allowing the system to fit within the room's steel grid. The floor that the table rests on in the plate stacks appears to be 5 foot thick solid concrete, which we suspect is sitting on bedrock.

### 3.4 Plate Holder

To assist with the plate holder design, we engaged a team of students at Worcester Polytechnic Institute (WPI) in Worcester, Massachusetts to help with the conceptualization, design, and manufacturing of the plate holding fixture. For the students this was a Masters Qualifying Project (MQP) which ran for the 2004-2005 school year.

Many previous digitizers had simple plate-holding mechanisms. Since the Schmidt plates were quite uniform and relatively thick, they were often held clamped on the edges with no other support. The large Schmidt plates reportedly sagged as much as 300 μm, which required changes of focus based on position[11].

The Harvard collection has about 25,000 plates of the 14 by 17 inch size with the bulk of the 500,000+ collection being the 8 by 10 inch size. The glass plates vary in thickness from about .060 inch (1.53 mm) to 0.15 inch (3.85 mm) with some having considerable wedge. In addition to the general thickness variation, the glass plates are sometimes crudely cut so that the sides are not straight or square, and the exposed emulsion varies in position on the plate. Approximately 5% of the plates in the collection have been repaired. These repaired plates are much thicker, with one or sometimes two pieces of clear glass used to hold the pieces together, and have many different kinds of tapes around the edges.

We required that it be easy for an operator to load and unload plates from the plate holder because it would by necessary to do this repetitively and quickly.

The final design has the normal, non-repaired, plates held emulsion side up, against a reference surface, supported from below by a platen with three layers of flash opal glass bonded together. The platen has a total thickness of about 0.45 inch (11.4 mm). The plates overlap the top layer of opal glass by 0.25 inch (6.35 mm) on all four sides. The plates are clamped 0.185 inch (4.7 mm) in from the each edge by pneumatic pressure pressing the plate from below against Viton® linear O ring material. This clamping action keeps the movable parts of the plate holder stable as the table accelerates and decelerates (~150 msec each) for each step.

The inner tray of the holder moves on a Bishop-Wisecarver Dual-Vee guide wheel and rail system. When the holder is in the loading and unloading position, the inner tray is extended out, like a CD/DVD player tray, onto a "catcher" rail system that is attached to the table. This gives easy access for loading and unloading the astrophotographs. When the tray is retracted into the holder, a pneumatic lift system automatically clamps the plates before the holder starts to move.

The entire digitizer was 3D modeled in Autodesk Inventor™. This program allows accurate estimations of the weight of each piece of the holder mechanism. By very carefully removing material not needed for structural strength from each piece of the holder, we were able to get its final weight to be 46.5 lbs (21 kg). From the 3D model we were able to prepare output for CNC manufacturing of the various parts that make up the holder and other pieces that we needed to build the overall machine.

### 3.5 Light system

The light system is comprised of 4 Lamina Ceramics Bl-31A0-1022 red (618 nm) LED arrays. These arrays each have 117 LEDs mounted as 39 clusters of 3 LEDs positioned over a 1 square inch area. The LEDs are mounted on a special co-fired ceramic and composite metal substrate that is thermally matched to the LED semiconductor material. The arrays are arranged electrically as two parallel sets of two arrays in series. Each series set is pulsed with a 14 volt, 5 amp constant current source, which was custom designed to fit within the light bar. This is 1/3 the rated current for the devices. At 1/3 the rated 2045 lumen output, the 4 arrays have an output of about 2600 lumens, the typical light output of an ordinary 150-watt light bulb  The arrays require ~120 Watts when active so they are somewhat more efficient than an incandescent light. However they are much smaller and easier to integrate into a very flat system and can be controlled more precisely with turn-on and turn-off times on the order of 4-5 ns. We use the LEDs both to provide light and to effectively act as a precisely controlled shutter. Software allows generation of 1μs to 50ms pulses. The pulse width to the LED arrays is adjusted for each plate during the digitizing process to match individual plate density variations (an initial calibration is made just at the plate center prior to the scan) and bring the exposure levels up to a near-constant mean (see section 4.2).

The LED arrays are temperature stabilized with a liquid-cooled copper heat sink. The LED duty cycle (exposure time / stage movement time) ranges from about 1%, to 10% so the average power to the light source is only a few watts. However, because the LED light output is a strong function of temperature, we wanted to be sure that the LED temperature remained constant. The liquid cooling system maintains the heat sink temperature to within a few tenths of a degree C of room temperature. We used a Coolermaster™ cooling system designed for water cooling CPU chips with a custom designed heat sink. This device monitors the temperature of the heat sink as well as the water reservoir.

### 3.6 Computer and Storage system

The computer is a dual 3.2 GHz Xeon system. The motherboard is a SuperMicro X5DAE with 2 Gigabytes of ECC 266 DDR memory. The video capture is done with a Dalsa (formerly Coreco) X64-CL Dual™ card. This card supports the two cameralink serial interfaces and has a 64 bit wide PCI-X interface to allow data transfers of up to 528 Mb/sec. The motion control card from Aerotech is a U500, which resides in a PCI slot. A Highpoint RocketRAID 2200, that supports 8 SATA II disks, is in another PCI-X slot and is used in a RAID 10 configuration to provide a mirrored and striped 1 Terabyte array (using 250 GB disks). The striping is needed to increase the write speed to disk, which is one of the potential system bottlenecks. All of the disks in the system are packaged in swappable containers. The system can run with both XP and Linux operating systems. The system is fully functional operating on the Microsoft Windows

XP operating system and nearly fully operational using a Linux operating system, which we are planning to move to for long term support. One of us (EJL) has developed the extensive code for DASCH for both Windows and Linux.

Our plan is to physically swap disks for bulk transfer of data from the digitizer to an online compute and storage facility (a 1 Gb fiber link does provide real time access for moderate sized data transfer, however). Once the disks are read into the online system, they will be archived as a non-rotating backup of the raw original data including the information needed to flat field the original data. The online system will hold the mosaic of each plate along with the WCS and photometric data. The online storage will also contain the catalog and the jpeg images of the logbook, plate jacket, and historical writing that was originally (before cleaning) on the back of the plate.

## 4. OPERATION

### 4.1 Focusing

Automatically focusing a telecentric lens system initially puzzled us. With both the image plane and object plane at infinity for a telecentric lens, images tend to be very consistent in size. As we experimented with the camera we first thought that using a Fourier transform would help us focus on the high frequency content of the photo emulsion (see Figure 2).

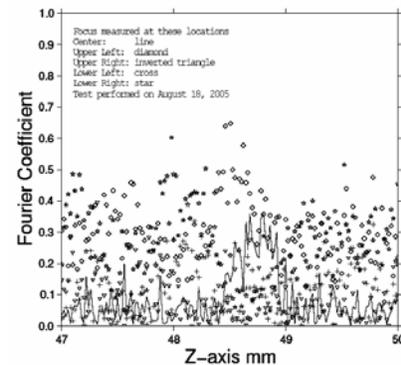

The focusing algorithm looks at 512 pixels taken from two diagonals in an X pattern across the CCD chip. The Fourier method turned out to be very noisy, and did not yield results that were easy to interpret (See Figure 4).

However as we experimented with the camera, the tool that we had written for looking at the distribution of pixels at a given A/D conversion unit (ADU) led us to observe that best focus occurred at the point of maximum black to white spread (see Figure 5). We now focus to the maximum standard deviation of the black and white areas of the image on the X diagonal pattern discussed above. This gives a good smooth indication of focus (see Figure 6). The algorithm works only for focusing on an emulsion surface.

Figure 4 Focus Measurement using Fourier Analysis

Focusing is done only occasionally since, except for repaired plates, all plates are clamped emulsion up on the plate holder and are therefore referenced to the same plane. The emulsion of repaired plates is usually sandwiched in the middle of two glass plates, placing the emulsion below the plate holder reference plane by the thickness of one or more of the clear glass support plates. So they require a special focusing.

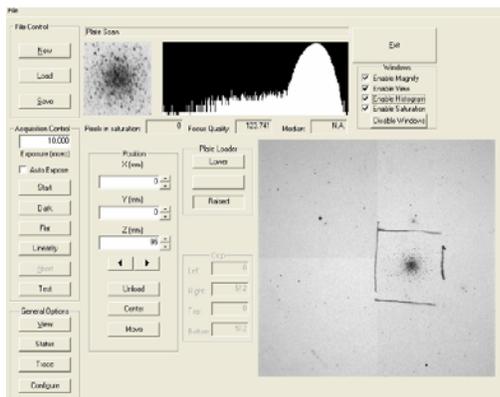

Figure 5 Image capture and histogram tool

### 4.2 Metering to adjust exposure times

Because the backgrounds of the plates vary from quite dark to very transparent, we need to automatically adjust the exposure time for each plate. The first thing done in the scanning process is to go to the center of each plate, measure the average background of a 512 by 512 pixel area and then calculate an exposure time for the plate. The exposure time is calculated to achieve an ADU count of 2710 for the center of the plate. This ADU count was selected experimentally to insure capture of the sky background pixel distribution while rejecting brighter emulsion defects. For very dark plates the exposure time is limited to 50 ms to prevent problems with hot pixels. Occasionally a plate will have a planet or very bright object (dark on the plates) where the metering occurs, and this causes problems. If the total number of saturated pixels in an individual exposure or the entire mosaic exceeds experimentally determined levels, then the operator will be immediately notified to rescan the plate with a manual exposure setting.

### 4.3 Flat fielding

When using a CCD for astronomical imaging, it is standard practice to image a uniform lighted object (or the twilight sky) and then create a pixel map of gain and offset needed to normalize to a "flat field". The sky is black and the star light fills the CCD wells. Dark currents accumulated over long exposure times can raise the black level of the sky.

For the digitizer, the problem is that the plates are photographic negative images of the sky so that the background is clear emulsion and the stars are black. Additionally, the clear emulsion (not really clear but frosted) is illuminated from below with a light source that is not perfectly uniform. Even if the light source were perfectly uniform, the edges of the plate holder will create non-uniform areas because of the shadowing of the light source. We attempt to compensate for all of these things and more by creating a flat field frame for each frame location in the whole plate area. To keep the optical path as close as possible to what it is with a plate in place, the flat field is done with a clear glass plate, nominally the same thickness as a normal photographic plate. For repaired plates we put two clear glasses in place of the repaired plate

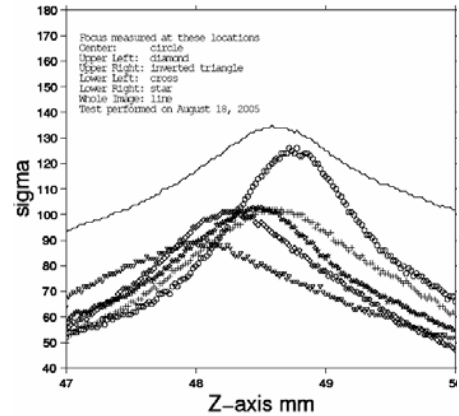

Figure 6  Focus Measurement using maximum Standard Deviation

The flat field images are obtained by taking 4 sets of 4 images at each frame location with the clear glass plate or plates rotated 180 degrees and flipped 180 degrees. The flat field is generated by first averaging the 4 exposures for each plate orientation and then taking the median of the 4 plate orientations. This eliminates effects from dust, finger prints, and other defects in the clear glass. The exposures are so short that there is no need for a dark current correction.

Flat field correction cannot compensate for any large areas of non-uniform photo density in the emulsion because that is on the plate itself. But our magnitude determination algorithms, which employ both local background measures and local calibration stars, address this issue.

### 4.4  Mosaic images

Once all the frames of the plate have been recorded, they must be assembled into a mosaic image of the whole plate. The file format for this image is FITS, a standard image format used in the astronomy community. When we create the mosaic at full resolution, we also make a smaller mosaic image with 16 by 16 binning (averaging a 16 x 16 pixel area into 1 pixel). The smaller image (~3 MB vs. 740 MB) is used as a starting point for the WCS fitting program, to speed up that initial calculation. A binned image is also converted to the JPEG format for use as a thumbnail for easy display in Web applications.

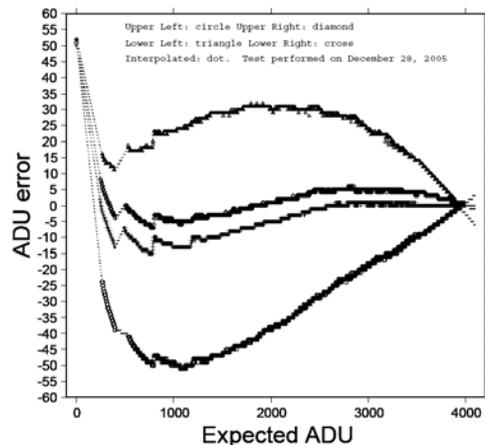

Figure 7 Nonlinearities of each CCD quadrant

The mosaic tool combines the individual frame tiles, the flat field frames for each location, and linearity results (discussed below) into full plate images. Options are available to control the binning of the images and to correct for rotation and magnification of the camera assembly with respect to the linear stage. Because the largest mosaics are 2.3 GB, the mosaic tool operates in a raster-scan mode to minimize use of CPU memory. If corrections for rotation and magnification are unnecessary, then the tool requires approximately 380 seconds under Linux to create a 2.3 GB mosaic. These full-size mosaics are necessary for photometric calibration, but final web-access utilities will allow smaller, full-resolution, plate extracts and binned mosaics.

When we first created mosaics, we found that the background white levels had a tiled pattern that was very obvious to the eye. We knew that single frames from the CCD chip had a quadrant pattern, but we expected the flat field algorithms to remove that in the mosaic. After much investigation, we realized that there were non-linear deviations in the CCD amplifier chain for each quadrant of the chip that were reflected in the Analog to Digital (A/D) conversion process. The differences introduced by the non-linear behavior would then be magnified by the flat field process and were a major contributor to the tiling effects. Figure 7 shows a plot of the error in ADU units for the measured output for each of the quadrants compared to the expected output as a light source is varied linearly. Note that this curve includes a correction for the CCD amplifier bias and obviates the need for a separate dark field exposure. Figure 8 shows a region of mosaic illustrating the tiling effect before and after correcting for quadrant non-linearities.

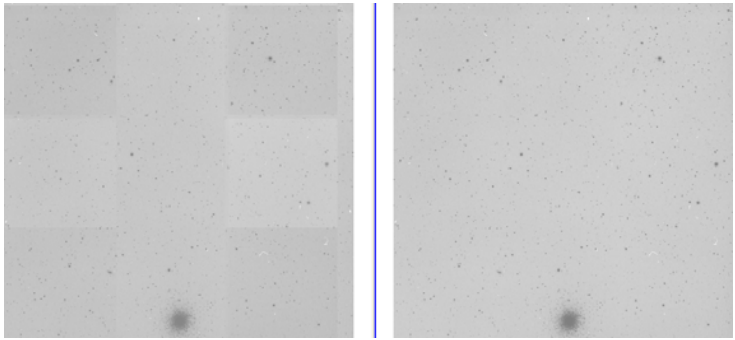

Figure 8 Before and after adjustment for different quadrant non-linearities. The "window-panes" on left are removed.

### 4.4.1 Image distortion

Image distortion was first measured using the x-y table with a single-mode fiber optic fiber (~6 μm diameter) illuminated with a green LED. This gave good results, but could not be interpolated well because the light was smaller than a pixel. We then moved to using a multimode fiber optic fiber of 62.5 μm diameter. The light from this fiber was large enough to give a good centroid to allow us to interpolate more accurately below the pixel resolution.

Figure 9 shows the results of comparing the difference in locations of a 64 by 64 square grid of table movements with the locations as read by the CCD chip. This shows that the lens is quite good in the center, with centroid locations of our artificial star agreeing with the specified table location to within ¼ pixel (~2.5 μm). At the outer corners of the CCD chip the centroids can disagree with the specified table locations by as much as one pixel. We are planning to investigate reducing these errors by even more careful alignment of the lens and CCD chip to the table. We have found the distortion to be a very sensitive indication of alignment of the optical axis, and of the working distances of the lens to the image plane and to the CCD plane.

### 4.4.2 Rotation

There are two sources of rotation in the final image. There is rotation of the CCD with respect to the axis of the table and the rotation of the plate with respect to the table. Both can cause rotations in the final image. However, the camera mount allows us to adjust the rotation of the camera with respect to the table to less than a pixel over the width of the CCD. Therefore, there is no need for rotation correction algorithms during the generation of the mosaic. The rotation of the plate in the fixture is corrected in the fitted WCS solution.

### 4.4.3 Magnification

Theoretically the telecentric lens has no magnification difference in or out of focus. In practice, lenses are not perfect and there will be some magnification difference from the design goal of unity and that will vary slightly with focus. We have found that the lens we have has a magnification of 0.99891 when in focus. The mosaic software takes this into account by using a projected pixel size instead of the actual CCD pixel size. This makes our effective resolution 2311.6 dpi.

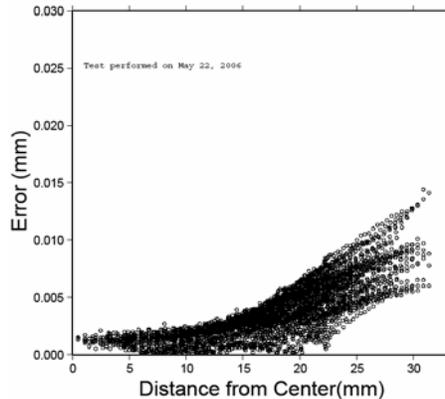

Figure 9 Measured errors from the lens

### 4.4.4 Scattered light

The CCD wells hold sufficient electrons and the camera electronics were designed to support true 12 bit (3.6 $D_p$) digitization. However, achieving this density range for the whole system can be difficult. The photo density range of the original black and white photographs is often considered to be in the range of 2.5 $D_p$ to 3.4 $D_p$. For these numbers we use photo density ($D_p$) as the range of tonal variations that the A/D conversion process can recognize, which seems to be the practice for scanning devices, rather than Optical Density ($D_o$) which measures light transmission. The photo density is measured by looking at the difference between the ADU readings of the white background and of the center of large dark objects. If there is a limit to the dynamic range in the system then as the light is increased, not only will the transparent areas of the photo increase in white ADU level but the dark area ADU of the photo will increase as well. We found that in our system, a large contributor to reduced range was light scatter off the interior walls of the lens and from the edges of the lenses themselves. After initial testing we had the manufacturer of the lens blacken the edges of the lenses and add optical flock paper to all of the internal sides of the lens to reduce scattering. The achieved photo density range for the digitizer (Log10 (measured black to white ADU range)) is about 3.55 $D_p$, very close to what is possible and greater than what is present in most all of the astronomical images that we will scan.

True Optical Density measurements are limited by shot noise in the CCD pixels and result in a 2.6 $D_o$ range for accurate transmission measurements.

After generating a number of mosaic images, we noticed small mirror images of areas near the edges of the plate a few tens of pixels wide separated by a clear line. We realized that vertical walls on the top part of the holder that clamps the plate, even though anodized black, were acting as mirrors. Adding black flock paper to those surfaces removed the artifacts.

## 5. METADATA

The card catalogs in use at the plate stacks, compiled from the logbooks, were all done by hand and so, unfortunately, are not readily machine readable. There has been an ongoing effort to convert that card catalog to an electronic format, but the process has been slow. After about 5 years with the efforts of only one part time worker, only 15-20% of the collection had been entered. Before we start production digitizing, we need to have the basic information about the plates (e.g. astronomical coordinates, exposure date and duration, etc.) in electronic format. To do that quickly requires copying the information into another format so that data entry can be done externally. We cannot send the historical logbooks or the card catalogs themselves.

After much consideration, we decided to go back and photograph all the pages of all of the original telescope logbooks. One of us (GC, an ATMoB member) investigated the resolution needed to capture the handwritten data and to start the imaging process. At this point we have determined that an ordinary 3 Mega pixel camera has enough resolution to provide very readable images after using image processing to enhance contrast. We have photographed about 20% of the logbooks at this writing. We also have acquired a page scanner for loose leaf logbooks to speed up the process for the part of the collection that is in that form. We are in the process of doing a trial run with an outside data entry source to make sure that what we are doing will work in a production environment. There are about 1200 logbooks, each with on average 100 pages and anywhere from 1 to 40 plates recorded per page.

Once we have the basic plate data entered, all of the image data can be referenced to the plate number. Each plate is stored in an archival paper jacket. The jacket has on it the pertinent plate information, often the same as the catalog. However there is sometimes additional information written by researchers as they have used the plate in the past. The plate itself may also have markings (on the back side, not the emulsion) in India ink or other kinds of colored inks to identify specific items of interest on the plate. For archival reasons we want to preserve these markings, but to enable the most accurate photometry, we need to clean these markings off the plate before the scan. To address these considerations, we first take a picture of the plate jacket (with a 10 Mega pixel Nikon D200 camera on a copy stand) and then for plates with any (historical) markings we also take a picture of the plate itself. The plate can then be cleaned and sent on to the digitizer

Studies of the optimal and fastest plate cleaning methods are in progress, including a semi-automated plate cleaner design by one of us (AS, an ATMoB member).

## CONCLUSIONS

We have developed a digitizer that meets the requirements we set out to achieve. Initial pilot runs, have shown handling and digitization rates of 50 plates per hour. WCS coordinates from catalogs have matched to measured centroids of star images across an entire plate to within a few pixels, even without any adjustment for original telescope distortions. And initial data indicates we can achieve photometric accuracy of $\leq .1$ magnitude from the plates.

We have developed and are continuing to refine the processes we believe necessary to capture all of the information on the plates and all of the information about the plates. We also have an analysis pipeline that can, often for the first time, look extensively at the information contained on each plate of Harvard's large collection. Our newly developed DASCH scanner is in the final stages of tuning and testing, and initial science projects have begun (on a small scale) as we next seek the sources of support needed to digitize the full Harvard collection.

## ACKNOWLEDGEMENTS

There are many who have contributed to this effort that we would like to acknowledge:
Photon Dynamics, Inc, who provided us with the camera;
F.J. Gray Inc., who provided us with much assistance with the glass platens;
Professor Kevin Rong of WPI and his students, who helped design the mechanical fixture to hold the plates and
William Weir of WPI's robotics lab, who provided much assistance in fabricating the parts for the plate holder;
Aerotech, Inc., who provided much assistance and guidance in the design;
Dave Monet, who early on provided much guidance and insight about requirements; and
Elizabeth Griffin, who encouraged us in every way to preserve and make the plate heritage more useful;
ATMoB, whose members have supported this project with their time and expertise.

We are grateful for the support of the National Science Foundation, through **NSF Grant AST-0407380** which has enabled the development of the DASCH project thus far.